\documentclass{PoS} 
\usepackage{graphicx}  
\usepackage{dcolumn}   
\usepackage{bm}        
\usepackage{amssymb}   

\usepackage{amsmath}

\usepackage{epstopdf}
\newcommand{\bee}{\begin{equation}}
\newcommand{\ee}{\end{equation}}
\newcommand{\beea}{\begin{eqnarray}}
\newcommand{\eea}{\end{eqnarray}}

\title{MCRG study of 12 fundamental flavors with mixed fundamental-adjoint gauge action}

\ShortTitle{MCRG study of 12 flavors}

\author{\speaker{Anna Hasenfratz}\\
        Department of Physics, University of Colorado, Boulder, CO-80309-390\\
        E-mail: \email{anna@eotvos.colorado.edu}}

\abstract{I discuss the infrared behavior of the SU(3) gauge model with  12 fundamental fermions.  Using a  Monte Carlo renormalization group technique I investigate the fixed point structure in the chiral limit and show that this system has an infrared fixed point and consequently conformal infrared dynamics. I am able to reach the FP   by using a new analysis method for the 2-lattice matching MCRG technique that   significantly reduces finite volume effects and by choosing a lattice action that avoids a spurious ultraviolet fixed point created by strong coupling lattice artifacts.}

\FullConference{ The XXIX International Symposium on Lattice Field Theory - Lattice 2011\\
July 10-16, 2011\\
Squaw Valley, Lake Tahoe, California}

\begin{document}
\section{Introduction}

Interest in strongly coupled gauge-fermion models has skyrocketed in recent years. Lattice methods offer  the most reliable tools, and in the Lattice 2011 Conference there were over two dozen contributed talks discussing these systems. Many of the talks considered  SU(3) gauge fields with 12 flavors of fundamental fermions, one of the most studied yet still  controversial  systems.  

2-loop perturbation theory predicts conformal behavior  with $N_f>8$ flavors, while summing up rainbow diagrams for a more reliable estimate puts the conformal window just below 12 flavors. Early numerical studies  used the Schrodinger functional approach with unimproved staggered fermion lattice action to numerically calculate the renormalization group $\beta$ function and concluded that the theory  has an infrared fixed point (IRFP)\cite{Appelquist:2007hu,Appelquist:2009ty}. Ref.\cite{Deuzeman:2009mh} considered the system at finite temperature and reached similar conclusion. On the other hand spectral quantities appeared to be more consistent with a chirally broken system\cite{ Fodor:2008hm,Fodor:2009wk, Jin:2009mc}. The first studies using Monte Carlo renormalization group (MCRG) techniques were not able to push deep enough into the strong coupling and remained inconclusive\cite{ Hasenfratz:2009kz,Hasenfratz:2010fi}. 

Recently a large scale study showed that an extensive  data set of spectral quantities prefers the chirally broken interpretation\cite{Fodor:2011tu}, but other groups interpreted the same data as compatible with  conformal behavior\cite{Appelquist:2011dp,DeGrand:2011cu}. 
In a recent work I revisited the 12 flavor SU(3) system using MCRG methods. Due to two improvements, one in the lattice action, the other in analyzing the MCRG data, I was able to cover a wider coupling range and  demonstrate that   the renormalization group $\beta$ function  develops a zero, signaling the existence of an infrared fixed point and  infrared conformality\cite{Hasenfratz:2011xn}. Other recent works include Refs. \cite{Ogawa:2011ki,Aoyama:2011ry}, reporting conformal infrared behavior based on the running of a renormalized coupling defined through twisted Polyakov loops.
Two other  works studied the phase diagram of the 12 flavor system \cite{Cheng:2011ic,Deuzeman:2011pa}. Both studies find three phases and indication of two bulk phase transitions, suggesting conformal behavior, but these works interpret  the nature of the phase transitions  quite differently.

Undoubtedly more work is needed to understand why different methods point to different conclusions and to find a consistent picture of the $N_f=12$ SU(3) system. The MCRG method is just one of the possible approaches to study the fixed point structure of the model, but the necessity to use an improved gauge action discussed in Ref.  \cite{Hasenfratz:2011xn} is quite general and applicable to all other investigations. In the following I summarize the results presented in Ref.  \cite{Hasenfratz:2011xn} and include some new results further supporting the conformal scenario of the $N_f=12$ model.

\section{Phase diagram of a system of 12 fermions with  fundamental-adjoint gauge action:}

The basic observation that makes the modified action necessary is the existence of an ultraviolet fixed point due to strong coupling lattice artifacts.  It is well known that the pure gauge SU(2) and SU(3) theories have a first order phase transition line in the fundamental and adjoint plaquette action space\cite{Bhanot:1981eb,Bhanot:1981pj}. This line ends in a second order point that corresponds to a (most likely trivial) ultraviolet fixed point (UVFP).  While the first order phase transition and the new UVFP are lattice artifacts and independent of the Gaussian FP and the continuum limit defined there, their existence can strongly influence, even completely change, the scaling behavior of the lattice model. 
 
 For example Ref. \cite{Hasenbusch:2004yq} studied the scaling of several observables of the SU(3) fundamental-adjoint pure gauge system and showed very large scaling violations, even lack of scaling, at couplings near the extension of the first order phase transition line. 
As expected, the scaling violations decrease with negative adjoint terms in the action, i.e. farther  from the second order endpoint.   
Refs. \cite{Tomboulis:2007re, Tomboulis:2007rn} studied the RG flow lines in the pure gauge SU(2) system and found that near the extension of the first order phase transition line along the fundamental plaquette action  the RG flows away from the UVFP-2 and turns   around sharply at negative adjoint coupling. This  again  indicates that in this region the system is strongly influenced by the FP associated with the second order phase transition.
In a recent work \cite{Hasenfratz:inprep} we studied the pure gauge fundamental-adjoint SU(2) system with the 2-lattice matching MCRG method.
 Our results show that near the first order line and its extension towards negative adjoint couplings the MCRG matching method breaks down, the system is no longer in the basin of attraction of the perturbative FP.

When two UVFPs exist numerical simulations have to stay in the vicinity of either one of them to describe the corresponding continuum physics.
The basin of attraction of the new UVFP  is at fairly strong coupling, and present day lattice simulations with 2 or 2+1 flavors are sufficiently far from it.    This, however, might not be the case in many fermion systems where interesting physics is expected to occur at strong gauge coupling. 

\begin{figure}
\vskip -.0cm
\begin{center}
\includegraphics[width=0.5\textwidth,clip]{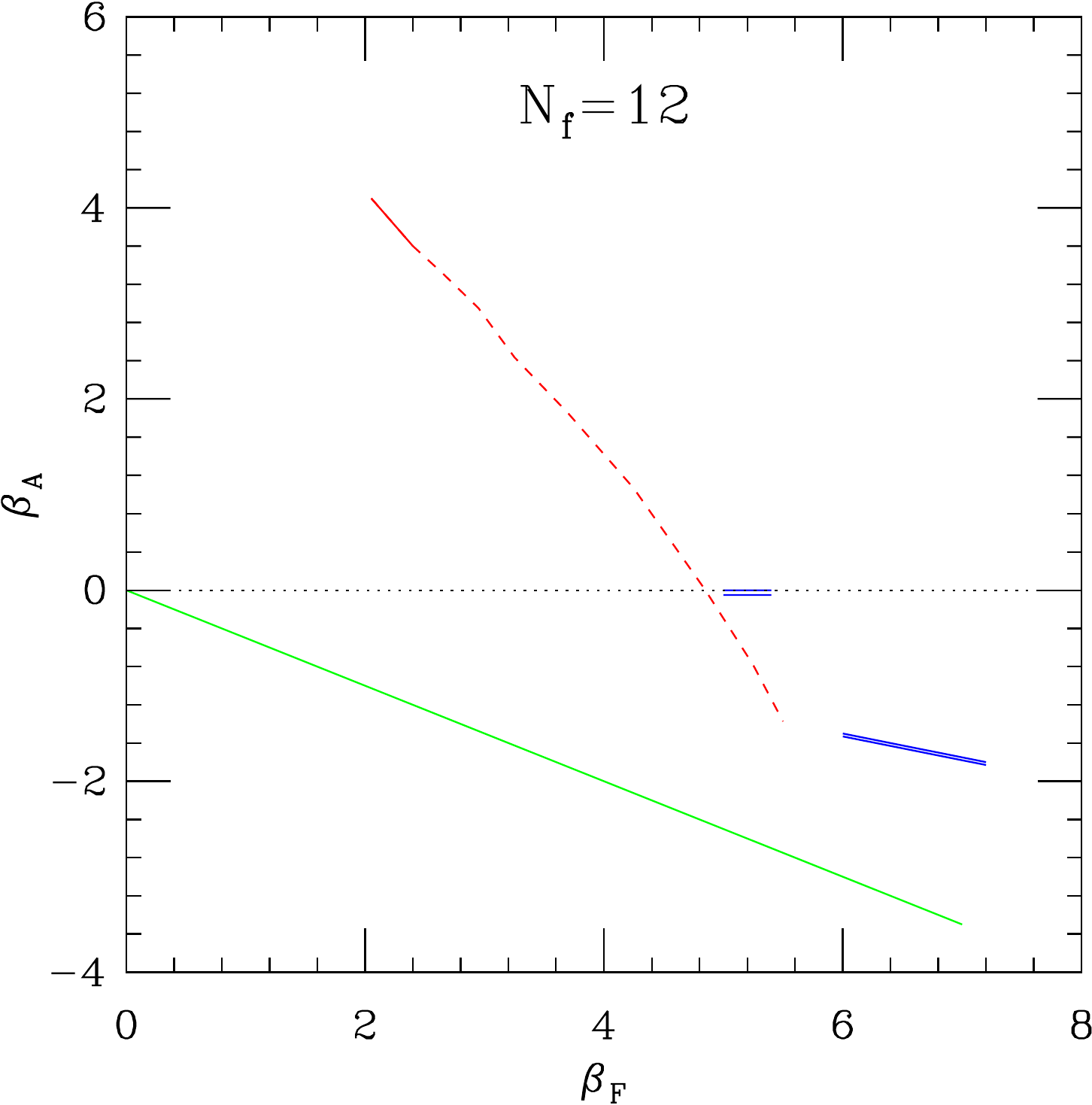}
\end{center}
\vskip -.1cm

\caption{ The red curve shows the approximate location of the phase transition /crossover in the fundamental-adjoint plane for the $N_f=12$ flavor system. 
} 

\label{fig:phase_transition}

\end{figure}

Large number of fermions could change the phase diagram of the pure gauge system so the first step is to study the phase diagram of the fundamental-adjoint plaquette gauge action with 12 fermions. I used nHYP smeared staggered fermions \cite{Hasenfratz:2007rf} and measured the plaquette, the specific heat through the derivative of the plaquette, the Polyakov line and the chiral susceptibility on $8^3\times4$, $8^4$, $12^3\times4$ and $12^3\times6$ lattices. The specific heat gave a very clear signal for a first order transition, continuing along a crossover line, as indicated by the solid and dashed red lines in Figure \ref{fig:phase_transition}. This phase transition/crossover has no dependence on the temporal lattice size, it is a bulk feature of the system. The 
 phase diagram of the 12 flavor system looks very similar to the pure gauge one. 
 
The horizontal blue line  in Figure \ref{fig:phase_transition} at $\beta_A=0$ indicates the region I studied in Ref. \cite{Hasenfratz:2010fi}. I found that matching became impossible at stronger couplings, to the left of the blue line,  most likely due to the nearby  crossover region. The leading order perturbative relation between the gauge coupling and the lattice couplings is
\bee
\frac{2N_c}{g^2}=\beta_F(1+2\frac{\beta_A}{\beta_F})\,.
\label{eq:beta_rel}
\ee
This suggests that the coupling $(\beta_F,\beta_A)$=(5.0,0.0) corresponds, at least perturbatively, to $(\beta_F,\beta_A)$=(10.0, -2.5). The latter point is quite far from the crossover along the $\beta_A/\beta_F=-0.25$ action line, indicated by the second blue line in Figure \ref{fig:phase_transition}. If the basin of attraction of the  perturbative FP is limited by the first order/crossover line and Eq. \ref{eq:beta_rel} is any indication of constant physics lines, than along  $\beta_A/\beta_F= -0.25$  one can reach considerably stronger couplings than with the $\beta_A=0$ fundamental action. 
Finally, the green line in the figure corresponds to $\beta_A/\beta_F=-0.50$, the limit where the adjoint plaquette overtakes the fundamental one and flips the system into a new universality class. 

Lattice simulations that want to investigate the properties of the Gaussian FP or the corresponding IRFP, if it exists, should stay away from the any spurious FP. This is true for any kind of investigations, be it Schroedinger functional or other running coupling calculations, spectral studies or  finite temperature investigations. 

\begin{figure}
\vskip -.0cm
\begin{center}
\includegraphics[width=0.5\textwidth,clip]{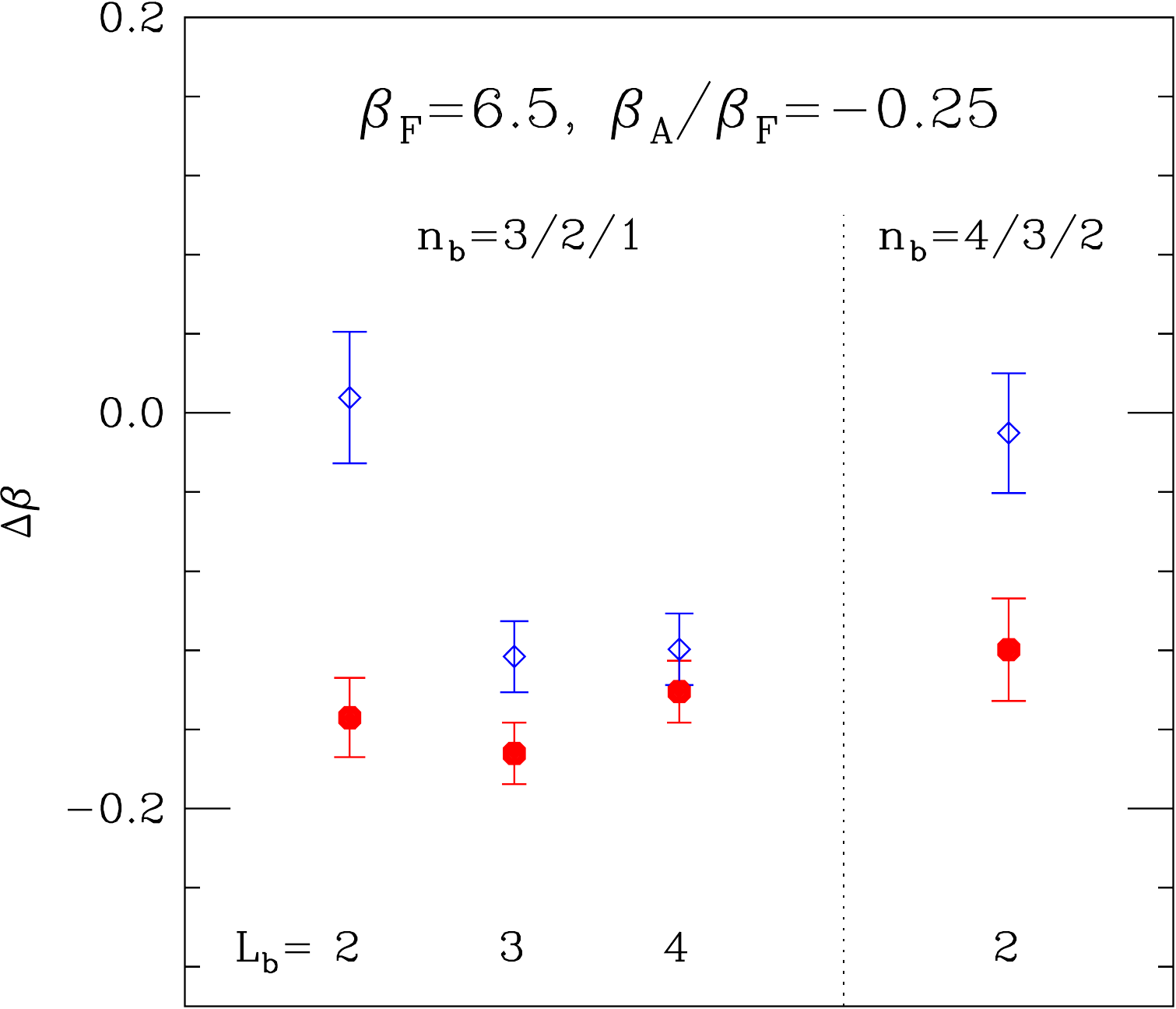}
\end{center}
\vskip -.1cm

\caption{ The optimized $\Delta\beta$ at $\beta_F=6.5$ with the action $\beta_A/\beta_F=-0.25$ . The red circles are the finite volume corrected predictions, the blue diamonds have no finite volume correction in the optimization step. The left side of the figure  shows results after comparing blocking steps $n_b=3/2/1$ on different volumes,  while right side is the result after one more blocking step. The data points are labeled by the final blocked lattice size $L_b$.  }

\label{fig:finite_vol_b65}

\end{figure}

\section{MCRG results with a fundamental-adjoint gauge action: }
The 2-lattice matching MCRG method is a powerful tool to numerically calculate the bare step scaling function, the discretized lattice analogue of the RG $\beta$ function \cite{Hasenfratz:2009kz,Hasenfratz:2010fi}. The method relies on  matching  observables after  RG blocking steps. Simulations do not have to be performed in volumes with lattice size  comparable or larger than the correlation length, as most of the finite size effects are cancelled by comparing blocked observables measured on identical blocked volumes. The success of the method  relies on the optimization of a blocking parameter that is set such that it assures the FP and its renormalized trajectory of the system is reached in a minimum number of blocking steps. It has not been recognized previously that this optimization procedure also introduces finite size effects. These effects, while much smaller than the finite size effects coming from  the matching of observables, can be important in systems where the step scaling function is very small. In this  section I describe how to minimize this secondary finite volume effects. 

Under repeated renormalization group blocking the action first flows towards the renormalized trajectory (RT), than along it, away from the UVFP. If the RG flow lines originating at $\beta$ and $\beta'$ hit the RT at the same points but require $n_b$ and $n_b-1$ blocking steps to do so,   the correlation lengths at $\beta$ and $\beta'$ differ by the scale of the block transformation, 2 in the present work. There are two steps necessary to achieve matching:
\begin{itemize}
\item {\it Matching:} Observables measured on $n_b$ times blocked configurations generated at $\beta$ have to match observables measured on $n_b-1$ times blocked configurations generated at $\beta'$ if the blocked actions are identical. Since blocking reduces the lattice size by a factor of two, in order to minimize finite size effects it is best to do the simulations on twice as large lattices at $\beta$ than at $\beta'$ so the final measurements are done on the same volume. The shift in the gauge coupling is defined as $\Delta\beta_{\cal O}(\beta;n_b,L_b) = \beta -\beta'$ if 
\bee
\langle  \mathcal O(\beta;n_b,L_b) \rangle = \langle  \mathcal O(\beta';n_b-1,L_b)\rangle\,,
\label{eq:match}
\ee
where $\langle \mathcal O \rangle$ denotes the expectation value of some short distance operator and $L_b$ is the volume after $n_b$ and $n_b-1$ blocking steps (the same for both sides of Eq. \ref{eq:match}).
\item {\it Optimization:} The quantity $\Delta\beta$ defined in the previous step can differ significantly from the step scaling function $s_b$ if the RG flow does not reach the RT in $n_b-1$ steps. Most RG block transformations have a free parameter, usually denoted by $\alpha$, that can be optimized to minimize the number of RG steps needed to reach the RT. The optimized parameter is defined as the one where consecutive blocking steps predict the same shift, 
\bee
\Delta\beta_{\mathcal O}(\beta;n_b,L_b,\alpha_{\rm opt})= \Delta\beta_{\mathcal O}(\beta;n_b-1,L_b,\alpha_{\rm opt})\,.
\label{eq:optim}
\ee 
Previous studies did not take the volume dependence of $\Delta\beta_{\mathcal O}$ into account and usually satisfied Eq. \ref{eq:optim} on  different volumes. The error introduced this way is small, but is important when $\Delta\beta_{\mathcal O}$ itself is small. Note that matching according to Eq. \ref{eq:optim} requires  simulations on three different volumes.
\end{itemize}

The procedure can be repeated with different $\mathcal O$ operators and the standard deviation between the predicted $\Delta\beta_{\mathcal O}$ values characterize the systematical errors of the matching. Results on larger volumes with more blocking levels provide further consistency checks.

Figure  \ref{fig:finite_vol_b65} illustrates the optimization. The simulations were done on volumes between $32^4$ and $4^4$  at $\beta=6.5$ with the $\beta_A/\beta_F =-0.25$ action. The RG blocking was based on 2 HYP smearing steps  and  5 different operators were used as described in Ref.\cite{Hasenfratz:2010fi}.

The left side of the figure shows the optimal $\Delta\beta$ with $n_b=3/2/1$ blockings and final blocked lattices of $L_b=$2, 3 and 4. The red octagons show the results of the optimized matching, a consistent value between all three volume series. In contrast the blue diamonds show the predicted $\Delta\beta$ without finite volume correction in the optimization. The result on the smallest volume set is clearly off, signaling a large finite volume effect. The two larger volumes predict consistent values, it appears that with my blocking transformations and 5 operators a finite volume of $L_b=3$ is already sufficient to minimize finite volume effects. The right side of the figure shows $\Delta\beta$ after one more blocking step, with $n_b=4/3/2$.  Again, the finite volume corrected optimized data is significantly different from the uncorrected one but both are consistent with the $L_b=2$ results of the left hand side. Comparing the finite volume corrected optimized results for  $\Delta\beta$ one can conclude that $\Delta\beta= - 0.13(2)$ on all three volume sequences and after $n_b=3/2/1$ and $n_b=4/3/2$ blocking levels. The negative value indicates that the RG $\beta$ function has crossed zero,  these measurements are on the strong coupling side of an IRFP.

\begin{figure}
\vskip -.0cm
\begin{center}
\includegraphics[width=1.0\textwidth,clip]{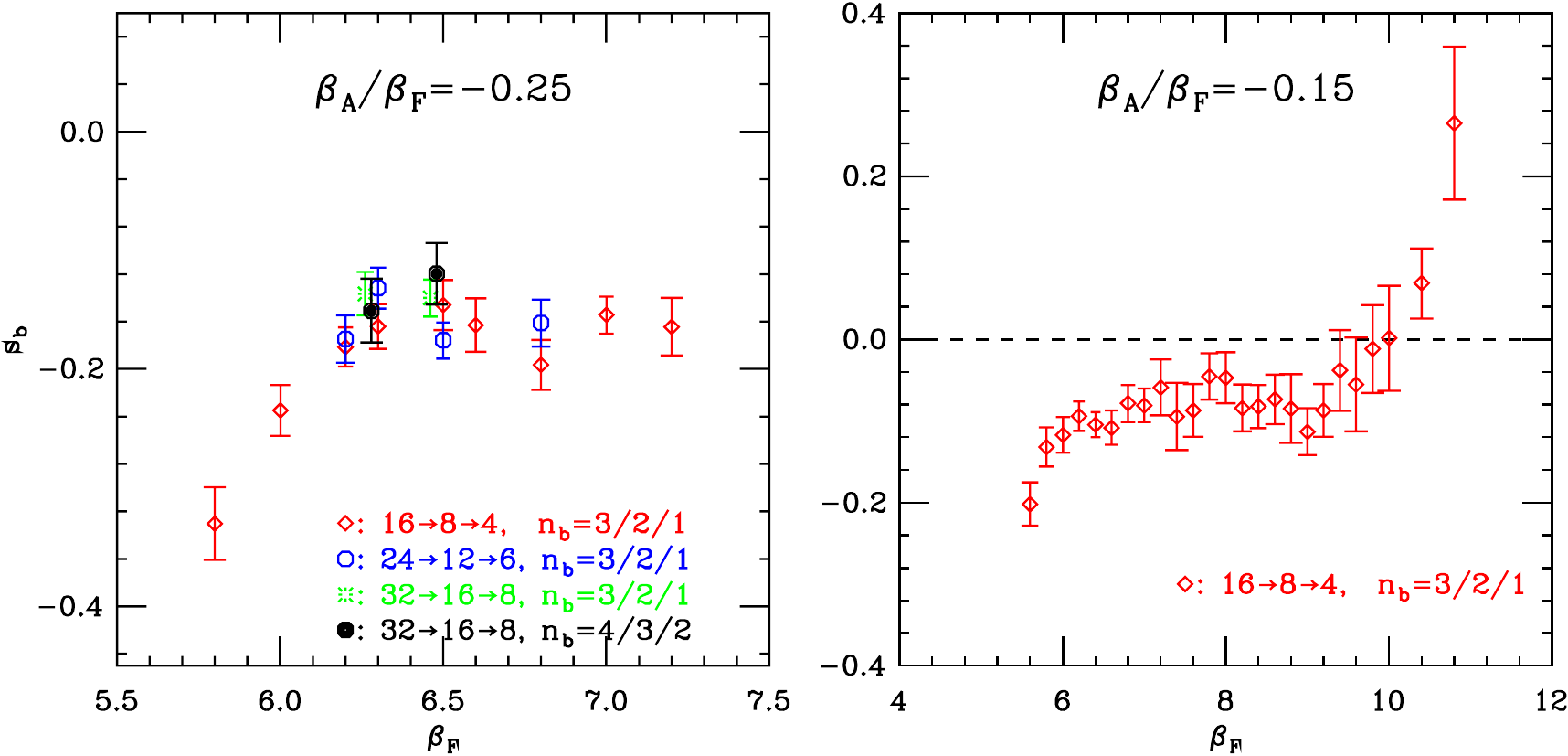}
\end{center}
\vskip -.1cm

\caption{ Left panel: the bare step scaling function for the  $\beta_A/\beta_F=-0.25$ action. The different symbols correspond to predictions from optimized matching on different lattice volumes and blocking levels. Right panel: the bare step scaling function for the $\beta_A/\beta_F=-0.15$ action. } 

\label{fig:db_all}

\end{figure}

The left panel of Figure \ref{fig:db_all} shows the result of optimized matching for the $\beta_A/\beta_F =-0.25$  action at several coupling values, volumes and blocking levels.  The data points are consistent and show a negative  $s_b$, consistent with the existence of an IRFP at some weaker coupling.  MCRG matching becomes increasingly sensitive to statistical errors at weaker coupling and I did not pursue matching any further with the $\beta_A/\beta_F=-0.25$ action. Rather I considered the action with $\beta_A/\beta_F=-0.15$. The result of MCRG matching using $n_b=3/2/1$ blocking is shown on the right panel of Figure \ref{fig:db_all}. This action has a zero in the step scaling function in the investigated region.

\section{Conclusion} 

I have investigated  the $N_f=12$ flavor SU(3) gauge model. This is a strongly coupled system and lattice studies have to control lattice artifacts, in particular the possible appearance  of spurious fixed points. I showed that the model with plaquette gauge action has such a spurious  UVFP in the fundamental-adjoint plane that can significantly influence lattice results. Adding a negative adjoint plaquette to the gauge action reduces this effect. Using a finite volume improved  2-lattice matching MCRG technique I was able to establish the existence of an IRFP for two actions with different fundamental-adjoint plaquette ratios, implying infrared conformality for this system.  

{\renewcommand{\baselinestretch}{0.86}
 \bibliography{lattice}
 \bibliographystyle{JHEP-2}}

\end{document}